\documentclass[prl,showpacs,superscriptaddress,twocolumn]{revtex4}

\usepackage{amsmath,bm}
\usepackage{graphicx}
\usepackage{epsfig}

\begin{document}

\newcommand{\vk}{{\vec k}}
\newcommand{\vK}{{\vec K}}
\newcommand{\vb}{{\vec b}}
\newcommand{{\vp}}{{\vec p}}
\newcommand{{\vq}}{{\vec q}}
\newcommand{\vQ}{{\vec Q}}
\newcommand{\vx}{{\vec x}}
\newcommand{\beq}{\begin{equation}}
\newcommand{\eeq}{\end{equation}}
\newcommand{\half}{{\textstyle \frac{1}{2}}}
\newcommand{\gton}{\stackrel{>}{\sim}}
\newcommand{\lton}{\mathrel{\lower.9ex \hbox{$\stackrel{\displaystyle<}{\sim}$}}}
\newcommand{\ee}{\end{equation}}
\newcommand{\ben}{\begin{enumerate}}
\newcommand{\een}{\end{enumerate}}
\newcommand{\bit}{\begin{itemize}}
\newcommand{\eit}{\end{itemize}}
\newcommand{\bc}{\begin{center}}
\newcommand{\ec}{\end{center}}
\newcommand{\bea}{\begin{eqnarray}}
\newcommand{\eea}{\end{eqnarray}}
\newcommand{\beqar}{\begin{eqnarray}}
\newcommand{\eeqar}[1]{\label{#1} \end{eqnarray}}
\newcommand{\pleft}{\stackrel{\leftarrow}{\partial}}
\newcommand{\pright}{\stackrel{\rightarrow}{\partial}}

\newcommand{\eq}[1]{Eq.~(\ref{#1})}
\newcommand{\fig}[1]{Fig.~\ref{#1}}
\newcommand{\eff}{ef\!f}
\newcommand{\alphas}{\alpha_s}

\renewcommand{\topfraction}{0.85}
\renewcommand{\textfraction}{0.1}
\renewcommand{\floatpagefraction}{0.75}

\title{ Jet tomography of high-energy nucleus-nucleus collisions 
 at  next-to-leading order }

\date{\today  \hspace{1ex}}

\author{Ivan Vitev}

\affiliation{Los Alamos National Laboratory, Theoretical Division,
MS B238, Los Alamos, NM 87545, USA}

\author{Ben-Wei Zhang}

\affiliation{Los Alamos National Laboratory, Theoretical Division,
MS B238, Los Alamos, NM 87545, USA}
\affiliation{Key Laboratory of Quark $\&$ Lepton Physics (Hua-Zhong Normal
University), Ministry of Education, China}

\begin{abstract}

We demonstrate that jet observables are highly sensitive to the 
characteristics of the vacuum and the in-medium QCD parton showers 
and propose techniques that exploit this sensitivity to constrain the 
mechanism of quark and gluon energy loss in strongly-interacting plasmas. 
As a first example, we calculate the inclusive jet cross section in 
high-energy nucleus-nucleus collisions to ${\cal O}(\alpha_s^3)$. 
Theoretical predictions for the medium-induced jet broadening and the
suppression of the jet production rate due to cold and hot nuclear 
matter  effects in Au+Au and Cu+Cu reactions at RHIC are presented. 

\end{abstract}


\maketitle

Hadronic jets~\cite{Sterman:1977wj}, collimated showers of energetic 
final-state particles, have long been regarded as the main tool to understand 
hard scattering  ($Q^2 \gg \Lambda_{QCD}^2$) processes in $e^++e^-$, 
semi-inclusive deeply inelastic scattering, and hadron-hadron collisions 
from first principles in Quantum Chromodynamics (QCD). Due to their large production 
rate~\cite{Sterman:1977wj}, jets are easily accessible by experiment 
and their landmark discovery has, arguably, stimulated some of the most important
developments in the perturbation theory of strong interactions. At present, 
cross sections for processes involving jets are  routinely calculated at 
next-to-leading order (NLO) and next-to-next-to-leading order results are also
becoming available~\cite{Campbell:2006wx}. This remarkable theoretical 
accuracy has  made precision jet studies  the optimal  method  for finding 
new physics beyond of the  Standard Model at very high energies~\cite{Olness:2009qd}.

An unprecedented possibility exists today to extend the theory of jets to 
energetic reactions
with large nuclei (A+A)~\cite{VWZ}. Experimental advances 
at this interface between particle and nuclear physics have already allowed 
proof-of-principle measurements of
jets at the Relativistic Heavy Ion Collider (RHIC)~\cite{JetExp}.
In this Letter we present the first calculation of jet cross sections and jet 
sub-structure for Au+Au and Cu+Cu reactions at RHIC that includes both
NLO perturbative effects and effects of the nuclear medium.
We demonstrate how jet observables can 
be used to gain insight into the mechanisms of parton interaction and 
energy loss~\cite{Wang:1991xy} in hot, dense, 
and  strongly-interacting quark-gluon plasmas (QGPs) that are created in such 
reactions. 
We determine quantitatively the relation between the in-medium modification and 
broadening of  parton showers in the QGP and the suppression in the observed 
cross section as a function of the jet cone radius $R$. 
Owing to their differential nature, jet observables will soon place stringent 
constraints~\cite{JetExp,jetlhc}  on the approximations and  
theoretical model assumptions that underlay the current competing approaches to 
quark and gluon energy loss in hot and dense QCD matter. Such discriminating power
cannot be achieved with measurements of leading particle quenching.  
This will, in turn, help eliminate the uncomfortably large 
systematic uncertainty in the extraction of the plasma properties, such as its 
density and transport  coefficients~\cite{Bass:2008rv}.


To take full advantage of jet physics in reactions with ultra-relativistic nuclei, 
calculations at 
${\cal O}(\alpha_s^3)$  are required~\cite{EKS}. In fact, 
the lowest order QGP-induced parton splitting  is also  manifested  in  experimental observables  
at the same  ${\cal O}(\alpha_s^3)$.  With the shortcut notation 
$d\{E_T, y,\phi\}_n =  \prod_{i=1,n}d y_{ i}  \prod_{j=2,n} E_{T_j} d\phi_j $  
for the independent final-state parton variables in the  
collinear factorization  approach,  
the simplest inclusive jet cross section at NLO can be expressed as follows~\cite{EKS}:
\begin{eqnarray}
\frac{d\sigma^{\rm jet}} {dE_Tdy}& =&\frac{1}{2!}
\int d\{E_T,y,\phi\}_2 
\frac{ d\sigma[2\rightarrow 2]}{d\{E_T,y,\phi\}_2}  S_2 (\{E_T,y,\phi\}_2) \nonumber \\
&& \hspace*{-1cm}+\frac{1}{3!}
\int d\{E_T,y,\phi\}_3   \frac{ d\sigma[2\rightarrow 3]}{d\{E_T,y,\phi\}_3 }
S_3 (\{E_T,y,\phi\}_3 ) \; . \qquad
\label{eq:CS_NLO}
\end{eqnarray}
Here, $E_{T\,i},y_i,\phi_i$ are the transverse energy, rapidity, and azimuthal angle
of the i-th particle ($i=1,2,3$), respectively, and
$\sigma[2\rightarrow 2]$, $\sigma[2\rightarrow 3]$ represent the production cross 
sections  with two and  three final-state partons. In Eq.~(\ref{eq:CS_NLO}) $S_2$, $S_3$
are phase space constraints and  $S_2 = \sum_{i=1}^{2} S(i)= 
\sum_{i=1}^{2} \delta(E_{T_i} - E_T)\delta(y_i -y)$
identifies the jet with its parent parton. Hence, only at NLO can  the 
dependence of the experimental observables on the jet cone radius $R$ and the jet finding 
algorithms~\cite{Campbell:2006wx,EKS}  be theoretically investigated. For an angular 
separation $R_{ij}=\sqrt{(y_i - y_j)^2 + (\phi_i - \phi_j)^2}$, defined for any
possible parton pair $(i,j)$,
\begin{eqnarray}
S_3 &=& \sum_{i<j} \delta(E_{T_i}+E_{T_j}-E_{T})
\delta\left(\frac{E_{T_i} y_i + E_{T_j} y_j}{E_{T_i}+E_{T_j}} - y\right)  \nonumber \\
&&\hspace*{-1.1cm} \times \theta\left(R_{ij} < R_{\rm rc}\right)
+  \sum_i S(i)\prod_{j\neq i} 
\theta\left(R_{ij} > 
\frac{(E_{T_i}+E_{T_j})R}{\max(E_{T_i},E_{T_j})} \right) \,. \;\; \quad
\label{eq:2to3}
\end{eqnarray}
In Eq.~(\ref{eq:2to3}) $R_{\rm rc} = \min\left(R_{sep}R, 
\frac{E_{T_i}+E_{T_j}}{\max(E_{T_i},E_{T_j})} R \right)$  determines when two partons 
should be  recombined in a jet. Here, $1 \leq R_{sep} \leq 2$ is introduced 
to take into account features of experimental cone algorithms, employed to improve 
infrared safety. Eq.~(\ref{eq:2to3}) establishes a correspondence between the
commonly used jet finders and the perturbative calculations 
to ${\cal O}(\alpha_s^3)$ with the goal of providing accurate predictions for  
comparison to data. For example,  $R_{sep} = 2$ yields a midpoint cone algorithm and  
$R_{sep} = 1$ corresponds to the 
$k_T$   algorithm~\cite{Campbell:2006wx,Seymour:1997kj}. 
$R$ is the cone size or parton   separation parameter, respectively.

\begin{figure}[t]
\begin{center}
\includegraphics[width=2.8in,height=2.9in,angle=0]{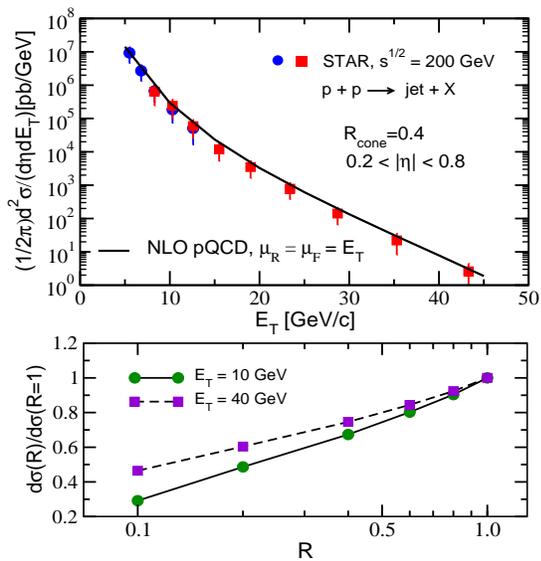} 
\vspace*{-.1in}
\caption{ Comparison of the NLO calculation to STAR experimental data~\cite{Abelev:2006uq} 
on the inclusive jet cross section for R=0.4 (top panel). The variation of the jet cross 
section with the cone size $R$ for  $E_T=10,\,40$~GeV around midrapidity at RHIC 
is also shown (bottom panel). }
\label{fig:STAR-pp}
\end{center}
\end{figure}
\begin{figure}[!b]
\begin{center}
\vspace*{-0.15in}
\includegraphics[width=3.in,angle=0]{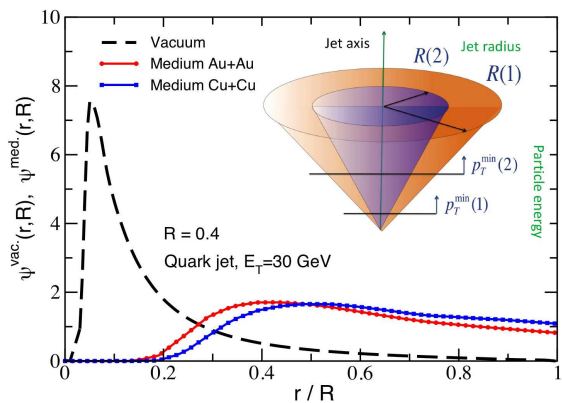} 
\vspace*{-.1in}
\caption{The differential jet shape in vacuum $\psi^{\rm vac.}(r,R)$ is contrasted to the  medium-induced 
contribution  $\psi^{\rm med.}(r,R)$  by a  $E_T = 30$~GeV quark in Au+Au and Cu+Cu collisions 
at $\sqrt{s_{NN}}=200$~GeV. The insert illustrates a method for studying the
characteristics of these parton showers.}
\label{fig:illust}
\end{center}
\end{figure}

In the top panel of Fig.~\ref{fig:STAR-pp} we show a comparison of the 
NLO calculation~\cite{EKS} of the inclusive jet cross section at ${\sqrt s}=200$~GeV p+p 
collisions  at RHIC to the STAR experimental measurement which uses a midpoint cone
algorithm~\cite{Abelev:2006uq} in the pseudorapidity range $ 0.2 \leq \eta \leq 0.8 $. 
Very good agreement between data and theory is achieved with a standard 
choice for the renormalization and factorization scales  
$\mu_R =\mu_f =E_T$. Variation of these scales within $(E_T/2,2E_T)$ leads to
less than ($+10\% , -20\%$) variation of the jet cross section. The bottom panel of  
Fig.~\ref{fig:STAR-pp} illustrates the significant dependence of 
$d\sigma^{\rm jet}/dydE_T$  on the cone size $R$, which, even in p+p reactions,
can exceed a factor of two. Analytically, the $\ln (R/R_0)$ scaling of the 
cross section can be understood from the $1/r$ angular behavior of the perturbative QCD 
splitting kernel.  Further insight into the underlying parton dynamics can be gained by 
examining the jet sub-structure, often characterized by the differential jet 
shape: 
\begin{equation}
 \psi(r,R) = \frac{d}{dr}\left\{\frac{\sum_i E_{T_i} \theta (r-R_{i, \rm {jet}})}
{ \sum_i  E_{T_i}  \theta(R-R_{i,\rm {jet}}) } \right\}\, . 
\label{eq:jetshape-1}
\end{equation}
In Eq.~(\ref{eq:jetshape-1})  $i$ stands for the sum over all particles 
in this jet and $ \int_0^R \psi(r,R)dr =1$.
Analytically, jet shapes can  be evaluated as follows~\cite{Seymour:1997kj,VWZ}:
\begin{eqnarray}
\psi^{\rm vac.}(r,R)&=&\psi_{\text{coll}}(r,R)\left( P_{\rm Sudakov}(r,R)-1\right) +
\psi_{\text{LO}}(r,R)  \nonumber \\
&& \hspace*{-2cm} + \psi_{i,\text{LO}}(r,R)  + \psi_{\text{PC}}(r,R)
+ \psi_{i,\text{PC}}(r,R) \;.
\label{totpsi}
\end{eqnarray} 
In Eq.~(\ref{totpsi})  the first term represents  the  contribution from the Sudakov-resummed 
small-angle  parton splitting; the second and third terms give the leading-order   
final-state and initial-state  contributions, respectively; the last two 
terms come from power corrections $\propto Q_0/E_T$, $Q_0 \simeq 2-3$~GeV, when one integrates 
over the Landau pole in the modified leading logarithmic  approximation (MLLA). 
This approach was shown to provide a very good description of the differential 
intra-jet energy  flow at the Tevatron~\cite{VWZ}, as measured by CDF II~\cite{Acosta:2005ix}. 
Thus, reliable  predictions for the jet sub-structure in p+p reactions at RHIC can be obtained 
and used as a baseline to study the distortion of jet shapes in more complex systems, such as p+A and A+A. 
An example of $\psi^{\rm vac.}(r,R=0.4)$ for a quark jet of $E_T = 30$~GeV is shown in Fig.~\ref{fig:illust}. 

When compared to a parton shower in the vacuum, the medium-induced quark and gluon 
splittings have noticeably different angular and lightcone momentum fraction 
dependencies~\cite{Vitev:2005yg,VWZ}. In particular, for energetic partons propagating 
in hot and dense QCD matter, 
the origin of the coherent suppression of their radiative energy loss, known as the 
Landau-Pomeranchuk-Migdal effect, can be traced to the cancellation of the collinear 
radiation  at $r < m_D / \langle\, \omega (m_D,\lambda_g, E_T) \, 
\rangle$~\cite{Vitev:2005yg}.  Here, the  Debye screening 
scale $m_D  = gT\sqrt{1+N_f/6}$ and $\langle \, \omega \, \rangle \simeq$~ few GeV. 
Thus, the medium-induced component of the jet, which is given by the properly normalized 
gluon bremsstrahlung intensity spectrum $ \psi^{\rm med}(r,R) \propto dI^{\rm rad}/d\omega dr $ 
within  the cone, has a characteristic large-angle distribution
away from the jet axis. This is illustrated in Fig. ~\ref{fig:illust} for central Au+Au and 
central Cu+Cu collisions at RHIC. We emphasize that accurate numerical simulations, taking into 
account the geometry of the heavy ion reaction, the longitudinal Bjorken expansion
of the QGP, and the constraints imposed by its experimentally measured 
entropy density per unit rapidity~\cite{VWZ}, have been performed for all physics results 
quoted in this Letter.

One can exploit the differences between the vacuum and the in-medium parton showers
by varying the cone radius $R$ ($R_{i,\rm jet} < R$) and a cut $p_T^{\min}$ ($E_{T_i} > p_T^{\min}$) 
for the particles "$i$'' that constitute the jet, to gain sensitivity to the properties  of 
the QGP and  of the mechanisms of parton energy loss in hot
and dense QCD matter. This is illustrated in the insert of Fig.~\ref{fig:illust}.  
The most easily accessible experimental feature of jet production in nuclear collisions 
is, arguably,  the suppression of the inclusive cross section in heavy ion reaction 
compared to the binary collision scaled, $\propto \langle  N_{\rm bin}  \rangle$, 
production rate in elementary nucleon-nucleon reactions~\cite{VWZ}: 
\begin{equation}
R_{AA}^{\text{jet}}(E_T; R,p_T^{\min}) =
\frac{ \frac{d\sigma^{AA}(E_T;R,p_T^{\min})}{dy d^2 E_T} }
{ \langle  N_{\rm bin}  \rangle
\frac{ d\sigma^{pp}(E_T;R,p_T^{\min})}{dy d^2 E_T}  } \; . 
\label{RAAjet}
\end{equation}
Eq.~(\ref{RAAjet}) defines a  two dimensional jet attenuation pattern 
versus $R$ and $p_T^{\min}$ for every fixed $E_T$. In contrast, for the same $E_T$, 
inclusive  particle quenching is represented by a  single 
value  related to the $R \rightarrow 0$ and
$p_T^{\min} \gg \langle \, \omega \, \rangle $ limit in Eq.~(\ref{RAAjet}).
Thus, jet observables are much more differential and, hence, 
immensely more powerful than leading particles and leading particle correlations 
in their ability to discriminate between the competing  
physics mechanisms of quark and gluon energy loss in dense QCD matter and between 
theoretical model approximations to parton dynamics in the QGP.

\begin{figure}[t]
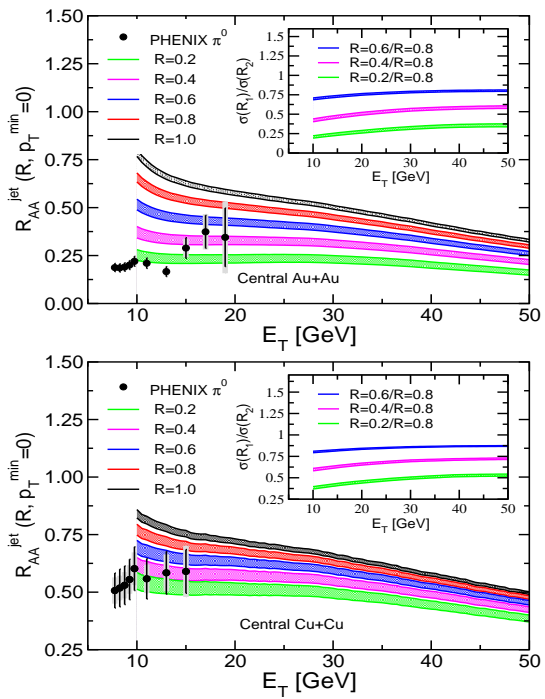

\begin{center}
\includegraphics[width=2.8in,height=1.8in,angle=0]{Raa-R-b3-jetAu-cold-IV.eps} \\
\includegraphics[width=2.8in,height=1.8in,angle=0]{Raa-R-b3-jetCu-cold-IV.eps}
\caption{ Transverse energy dependent nuclear modification factor $R_{AA}^{\rm jet}$
for different cone radii $R$ in $b=3$~fm Au+Au (top panel) and
Cu+Cu (bottom panel) collisions at $\sqrt{s_{NN}}=200$~GeV. Inserts show ratios of jet cross 
sections for different $R$ in nuclear reactions versus  $E_T$. }
\label{fig:Raa-jet}
\end{center}
\end{figure}


We now focus on the first complete theoretical result at NLO for 
$R_{AA}^{\text{jet}}$ versus the jet cone size $R$ for 
Au+Au and Cu+Cu collisions at RHIC. We use the reaction operator approach 
to non-Abelian energy loss~\cite{Gyulassy:2000fs}, in the limit of weak 
coupling between the jet and the plasma with $\alpha_s \sim 0.3$,  to evaluate  the probability 
distribution $P_{q,g}(\epsilon,E)$ that quarks and gluons, respectively, will
lose a fraction of their energy  $\epsilon = \sum_i \omega_i/E$ due to medium-induced 
bremsstrahlung.  Next, we determine the fraction of this energy that will 
be redistributed inside the jet:
\begin{eqnarray}
f_{q,g} \equiv f(R,p_T^{\min})_{q,g}= \frac{\int_0^R dr \int_{p_T^{\min}}^{E_T} d\omega \, 
\frac{dI^{\rm rad}_{q,g}}{d\omega  dr }}
{\int_0^{R^{\infty}} dr \int_{0}^{E_T} d\omega \, \frac{dI^{\rm rad}_{q,g}}{d \omega dr} } \; .
\end{eqnarray}
While such redistribution may affect the jet shape, it will not affect the jet cross 
section. For example, when  $R \rightarrow R^\infty$ and $p_T^{\min} \rightarrow 0$ ($f_{q,g} = 1$)
final-state QGP-induced effects to inclusive or tagged jet cross sections vanish. 
Parton interactions in the strongly-interacting plasma, however, are not the only 
many-body QCD effects that will alter the measured jet cross section.  
Cold nuclear matter (CNM) effects prior to the QGP formation~\cite{Vitev:2008vk}
must also be included in accurate theoretical
calculations of hard probes production in nuclear collisions and we first 
evaluate $\frac{d\sigma^{\rm CNM, NLO}} {d^2E_Tdy} $. We find that in the 
kinematic region of interest, $10 \; {\rm GeV} \leq E_T \leq 50$~GeV around
midrapidity at RHIC $\sqrt{s_{NN}} = 200$~GeV collisions, the EMC effect and  initial-state
energy loss~\cite{Vitev:2008vk} play a dominant role. Next, we determine the
relative fractions $n_{q,g}$ of quark and gluon jets in this inclusive cross 
section $(n_q+n_g=1)$. These are well defined at leading order~\cite{Dasgupta:2007wa}
and separation of the inclusive cross section into $\frac{d\sigma^{\rm CNM, NLO}_{q,g}} {d^2E_Tdy} $
is necessary to properly describe parton energy loss in the QGP,
which scales with the quadratic Casimir in the  corresponding  representation of SU(3)  
($C_A/C_F = 2.25$).  At NLO there exists an ambiguity of  ${\cal O}(\alpha_s)$ in this 
separation~\cite{Dasgupta:2007wa}, which has a very small
effect on inclusive jet observables.

We calculate the medium-modified jet cross section per
binary nucleon-nucleon scattering as follows ($p_T^{\min}=0$):
\begin{eqnarray}
\frac{1}{\langle  N_{\rm bin}  \rangle} 
\frac{d\sigma^{AA}(R)}{d^2E_Tdy} &=& \int_{\epsilon=0}^1
d\epsilon \; \sum_{q,g}  P_{q,g}(\epsilon,E) \nonumber \\
&& \hspace*{-1cm}\times
\frac{1}{ (1 - (1-f_{q,g}) \cdot \epsilon)^2} 
 \frac{d\sigma^{\rm CNM,NLO}_{q,g}(R)} {d^2E^\prime_Tdy} \; . \qquad
\label{eq:JCS-AA}
\end{eqnarray}
In Eq.~(\ref{eq:JCS-AA})   
$(1-f_{q,g}) \cdot \epsilon$  represents the fraction of the 
energy of the parent parton that the medium re-distributes outside of the cone of
radius $R$. The measured cross section is then a probabilistic superposition of 
the  cross sections of protojets of initially larger energy
$E^\prime_T = E_T / (1 - (1-f_{q,g})\cdot \epsilon)$. Our results for
the nuclear modification factor of inclusive jets $R_{AA}^{\rm jet}$ in central Au+Au
and Cu+Cu collisions with $\sqrt{s_{NN}}=200$~GeV at RHIC are presented 
in Fig.~\ref{fig:Raa-jet}. Each band illustrates a calculation for a  
$\sim 20\%$ increase in the rate of parton energy loss (lower bound) relative 
to our default 
simulation (upper bound). Experimental data on leading $\pi^0$ suppression for these
reactions is only included for reference. 
A continuous variation of $R_{AA}^{\rm jet}$ with the cone radius $R$ is 
clearly observed in  Fig.~\ref{fig:Raa-jet} and shows the sensitivity of the inclusive 
jet cross section  in high-energy nuclear collisions to the characteristics of
QGP-induced parton shower. For $R \leq 0.2$ the quenching of jets approximates the
already observed suppression in the production  rate of inclusive high-$p_T$ particles.
It should be noted that in our theoretical calculation CNM effects contribute close 
to $1/2$ of the observed attenuation for $E_T  \geq 30$~GeV. These can be 
drastically reduced at all $E_T$ by taking the ratio of two differential cross 
section  measurements for different cone radii $R_1$ and $R_2$. 
A few selected examples are shown in
the inserts of  Fig.~\ref{fig:Raa-jet}. 

\begin{table}[!tb]
\begin{tabular}{|c|c|c|c|c|}
\hline
  \ \ \ \ \  $ \langle \, r/R \, \rangle $ \ \ \ \ \  &  \ \ Vacuum  \ \  
&  \ \ Medium  \ \  & \ \ Total \ \  & \ \ \ \ \ $\Delta$ \ \ \ \ \   \\
\hline
    Au+Au                 &  0.271   & 0.601   &  0.283  & 4\%    \\
\hline
    Cu+Cu                 &  0.271   & 0.640   &  0.272 &  0.4\%   \\
\hline
\end{tabular}
\caption{ Mean relative jet radii $\langle \, r/R \, \rangle$
in the vacuum, for complete parton energy loss in the medium, and for the 
realistic case of jets in Au+Au and Cu+Cu collisions at RHIC. 
We considered a radius $R=0.4$ and  transverse
energy $E_T = 30$~GeV at $ \sqrt{s_{NN}}=200$~GeV. The fractional
QGP-induced broadening $\Delta\langle \, r/R \, \rangle $ 
is also shown.}
\label{table:mean-radii}
\end{table}

Inclusive jet cross sections and jet shapes in nuclear collisions are closely 
related~\cite{VWZ}:
\begin{eqnarray}
&& \psi_{\rm tot.}\left(\frac{r}{R}\right) =  \int_{\epsilon=0}^1
d\epsilon \; \sum_{q,g} \frac{P_{q,g}(\epsilon,E) \, \chi_{q,g}(R;E_T,E_T^\prime) } 
{ (1 - (1-f_{q,g}) \cdot \epsilon)^3} \,  \nonumber \\
&& \times  \Bigg[ (1- \epsilon) \;
\psi_{\rm vac.}^{q,g}\left(\frac{r}{R};E^\prime \right) + f_{q,g}\cdot \epsilon \;
\psi_{\rm med.}^{q,g}\left(\frac{r}{R};E^\prime \right) \Bigg] \; ,  \qquad
\label{psitotmed}
\end{eqnarray}
 $\chi_{q,g}(R;E_T,E_T^\prime) /  { \langle  N_{\rm bin}  \rangle } 
=   \frac{d\sigma^{\rm CNM,NLO}_{q,g}(R, E_T^\prime )}{d^2E_T^\prime dy}    \Big/ \frac{d\sigma^{AA}(R)}{d^2E_Tdy}$. 
It should be noted that 
vacuum and medium-induced parton showers become more collimated with increasing 
$E_T^\prime$ and the mean relative jet width
$ \langle r/R \rangle = \int_0^1 d(r/R) (r/R) \psi\left({r}/{R}\right)\ $ is 
reduced~\cite{VWZ}. Consequently,
the striking suppression pattern for jets, shown in Fig.~(\ref{fig:Raa-jet}), can 
be accompanied by a very modest growth in the observed 
$\langle r/R \rangle_{\rm tot.}$.
We show in Table~\ref{table:mean-radii} the relative widths in the vacuum, for a
hypothetical case  of complete parton energy loss ($P_{q,g}(\epsilon)=\delta(1-\epsilon)$) 
that falls inside of $R$   ($f_{q,g}=1$), and for our realistic simulation 
of  $E_T = 30$~GeV jets of $ R=0.4$ 
in central  Au+Au and Cu+Cu collisions at RHIC. We find that, on average,  
jet broadening  $\Delta\langle r/R \rangle = 
(\langle r/R \rangle_{\rm tot.} - \langle r/R \rangle_{\rm vac.}) 
/\langle r/R \rangle_{\rm vac.}$  is  $ <5\%$. Larger effects are expected 
near the core $r\rightarrow 0$ and the periphery
$r\rightarrow R$ of the jet~\cite{VWZ}.

In summary, the principle goal of this Letter is to bring into focus 
the immense possibilities 
that jets offer as tomographic probes of the QGP created in ultra-relativistic 
nuclear collisions.  To demonstrate the unprecedented sensitivity of jet observables 
to the characteristics of the 
vacuum and the medium-induced parton showers, we presented first results for 
the related cross sections 
and  shapes as a function of the cone radius $R$ at next-to-leading order 
${\cal O}(\alpha_s^3)$ 
in p+p, central Au+Au, and central Cu+Cu collisions at RHIC. 
Our theoretical predictions include a 
most detailed account of cold and hot nuclear matter effects on jet production and distortion 
in heavy ion reactions and are of immediate relevance to upcoming PHENIX and STAR experimental 
measurements. We fully expect that this work will inspire future studies of 
inclusive and tagged jets, 
jet sub-structure and event shape observables. In their entirety, such studies
will provide  first-principles insights into the many-body QCD parton dynamics at 
ultra-relativistic energies and shed light on the relative importance of collisional versus 
radiative  energy loss and on the applicability of weak versus strong jet-medium coupling regimes.

\vspace*{-.05cm}

{\bf Acknowledgments:} We thank D.~E.~Soper and F.~I.~Olness for helpful discussions. 

\end{document}